# Two simple experiments using FFT with digital devices in the introductory physics laboratory

Eugenio Tufino, Luigi Gratton and Stefano Oss

Physics Department, University of Trento, 38123 Povo (Trento), Italy

*Abstract*

We propose two experiments suited for high school and/or undergraduate physics laboratory which are aimed to the discovery of the practical meaning and usefulness of the FFT analysis as a mathematical and graphical instrument, intended to discriminate different periodic signals in various situations, ranging from basic, demonstrative measurements to very complex, real-research experiments. In our proposal, quite inexpensive digital instruments can be used, yet with accurate out-of-the-box results.

**1. Introduction**

As it is well known, the Fourier transform is widely used in various fields of science, including engineering, physics, computer science and mathematics. It allows the spectral analysis of periodic functions of any kind. The use of this mathematical construct has become even more pervasive and effective thanks to the Fast Fourier Transform (FFT) algorithm, which allows a much more efficient computation of the Discrete Fourier Transform (DFT) of a signal. Quite obviously, several applications of the FFT technique concern oscillations and waves, of whatever nature, such as light and sound [1]. As a consequence, a basically unlimited number of case studies opens in principle at almost every education level. For example, it offers the opportunity to introduce the use of FFT in experimental physics activities even in high school introductory physics courses [2,3]. The main idea is that students can explore waves in both frequency and time domains to obtain complementary information for multiple, relevant questions concerning the proper modeling of numerous physical phenomena. In quite recent years, FFT can be accomplished with low-cost devices, such as microcontrollers (e.g., Arduino) as well as several free apps to be used with a smartphone (for example Phyphox, Physics toolbox, etc.). There exist sensor-equipped devices, such as iOLab and PASCO units, whose graphical interfaces allow the real-time execution of FFT. It is also possible to use programming language libraries (e.g., Python, Matlab, etc.) that provide FFT functionality for data acquired with the above mentioned low-cost digital devices. Students can thus be directly introduced to the FFT technique which, as we show in this work, provides, also in terms of graphical representations, useful information on periodic phenomena with a valuable insight in the frequency domain [4].

In this paper we propose two laboratory activities, suited for introductory physics courses at various levels, devoted to illustrate how FFT operates. As a measuring device, we adopt an iOLab unit [5]. In the first activity, we measure with a photo detector the light signal coming from a simple, low-cost Michelson interferometer with the superimposition of the sound generated by a speaker placed near to the apparatus. This activity represents an oversimplification of contemporary gravitational wave research, in which any extraneous signal has to be identified and eventually filtered out [6]. In the second activity, the analysis of frequency of AC current flowing in the home power grid is carried out. As it will be discussed in the following, this measurement brings to the attention of students the relevance of applying an anti-aliasing procedure to the obtained signal, as it quite often happens in DFT processes.

2. **A noisy experiment with a Michelson interferometer**

Besides its great historical significance for well-known reasons, the Michelson interferometer continues to be used in many laboratories for various measurements, one of the most famous being the detection of gravitational waves, an experiment which was awarded by the 2017 physics Nobel Prize [7]. Here, we propose an educational rendition of this same experiment based on a low cost tabletop interferometer made and distributed by Nikhef for educational purposes [8], even if other similar apparatuses are available [9-11]. The basic implementation of the setup is schematized, along with its practical realization, in Fig.1. To complete the Nikhef interferometer, which includes the beam splitter, the mirrors and a red laser, a photo detector is needed: here, we used the sensor onboard of an iOLab unit (working at 12 bit with a maximum 4.8 kHz sample rate).

In our experiment, we introduce some periodic signal to be detected having influenced the regular functioning of the apparatus. At this aim, we placed a loudspeaker somewhere on the floor in the vicinity of the interferometer. The speaker was driven by a sinusoidal function generator at a given, arbitrary frequency in the audible range. The mechanical vibrations associated with the acoustical signal affect the interference pattern produced by the interferometer exactly in the same fashion as it happens in the LIGO-Virgo apparatuses under the action of a gravitational wave. Here, one can extract the frequency of the loudspeaker by performing a Fourier analysis of the output signal. This approach will of course allow the detection of any periodic event: in this way, the apparatus can be used to isolate and eventually remove unwanted contributions.

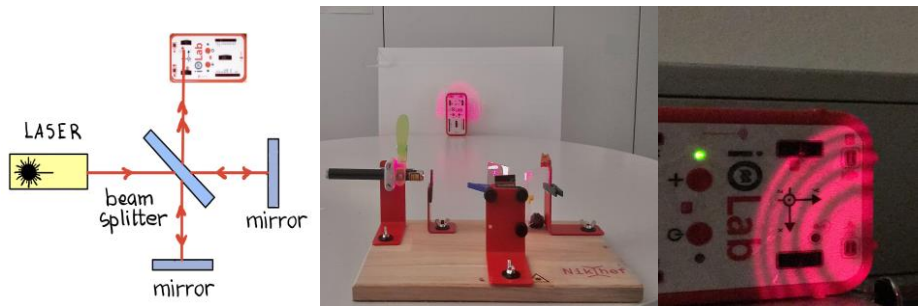

Fig. 1 Basic, typical layout of the Michelson-Morley interferometer in which the iOLab unit works as a light sensor (left); Nikhef interferometer and iOLab (center); iOLab light sensor and interference pattern (right).

The experiment is done according to quite a standard procedure in which the interferometer mirrors are adjusted so as to have the expected interference pattern on the screen, placed at about 50 cm distance. Here, the light sensor is placed in approximated correspondence with the first dark fringe. By using the iOLab sensor and its software interface for data analysis, the light signal is measured and analyzed: one can directly apply the FFT to obtain a representation in real time in the frequency domain. So, the interesting part of the experiment, as said above, occurs when there is a periodic contribution to the output signal since it will be immediately recognized usually as single peaks in the Fourier transformed signal. We show in Fig.2 a typical outcome of this procedure: we notice a peak at the frequency of the loudspeaker which was carrying a sinusoidal sound at about 137 Hz. In order to properly acquire and obtain a DFT, the analyzing device operates in a given, predetermined window of time and special attention is needed to avoid spurious oscillations in the final signal (this procedure is known short-time Fourier transform, see [1]). This is automatically done with the iOLab software: Fig.2 also depicts the moving window (colored in light grey) in which the signal is acquired. It is an advantageous practice to restrict as much as possible this window centered about the signal of interest in order to maximize the signal-to-noise ratio in the frequency domain (once again, this procedure is easily done with the iOLab software). Should one use a different system to analyze the experimental output, such as an Arduino based device, several freely available codes - also as python Jupyter notebooks - implementing the selectable time-window can be easily found and used, see for example [12], which contains a demonstrative code written explicitly for the present work.

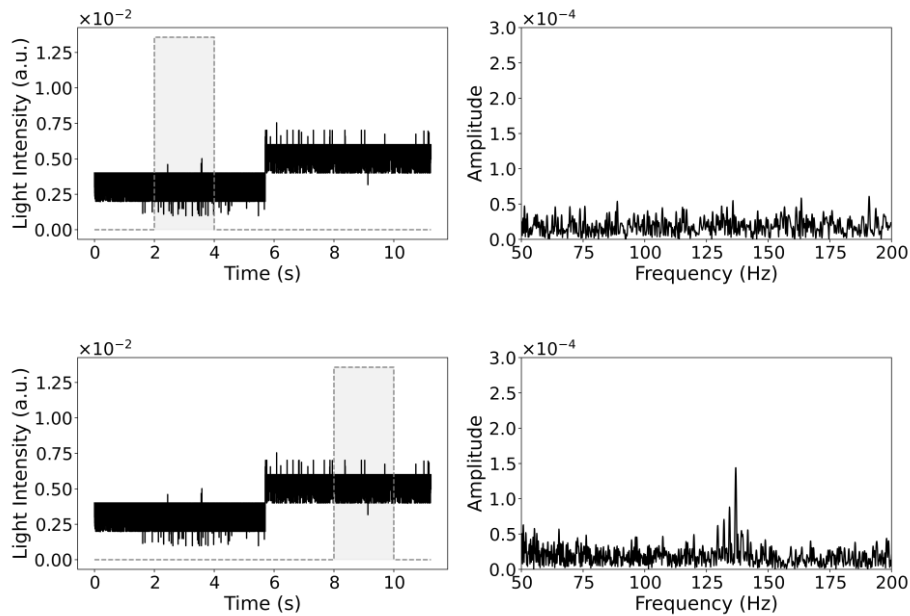

Fig.2 iOLab light sensor signals (left) and their FFT corresponding to the selected time window (right). In the top row, the window (in light grey) takes an interval of time without any external disturbance. In the bottom row, the FFT is performed with the loudspeaker switched on and the corresponding peak at about 137 Hz is clearly visible. The jump of the light signal at about 6 seconds is caused by the onset of mechanical vibrations of the interferometer due to the loudspeaker and, as a consequence, of the deviation of the fringes as measured by the iOLab sensor.

In Fig.3 we show a FFT in which two peaks are visible: besides the acoustic contribution representing the vibrating loudspeaker - which is, as already said, the counterpart of the gravitational wave - a second peak is located at 100 Hz. This peak is easily shown to be caused by the oscillations of the ambient light in the laboratory since it is provided by a lamp powered with a 50 Hz AC voltage (in the majority of the world, the frequency of power lines is 50 Hz, while in North America it is 60 Hz): the light pulsates at twice this frequency because of the half-rectification of the alternating current power supply. This could be in its own right an interesting aspect to be addressed in the school lab: the light intensity depends on the square of the electrical current, i.e. the frequency of the intensity will be twice than that of current amplitude. We mention that a similar measurement of the ambient light frequency can be taken, using a slightly more time-consuming procedure, through video analysis of the bulb video captured in slow-motion mode with a smartphone camera [13]. As already pointed out, the experiment with the educational interferometer can somehow be seen as a simplification of the actual situation in the large gravitational wave interferometric labs, where extreme care is taken in the discrimination of unwanted, yet present and unavoidable, spurious oscillations, also known as "instrumental effects": in our simulation, the ambient light signal at 100 Hz can in fact be taken as an example of disturbance which is promptly extracted from the overall signal, while the requested contribution (here the loudspeaker sound at 137 Hz) is left in clear evidence and eventually distinguished from other sources.

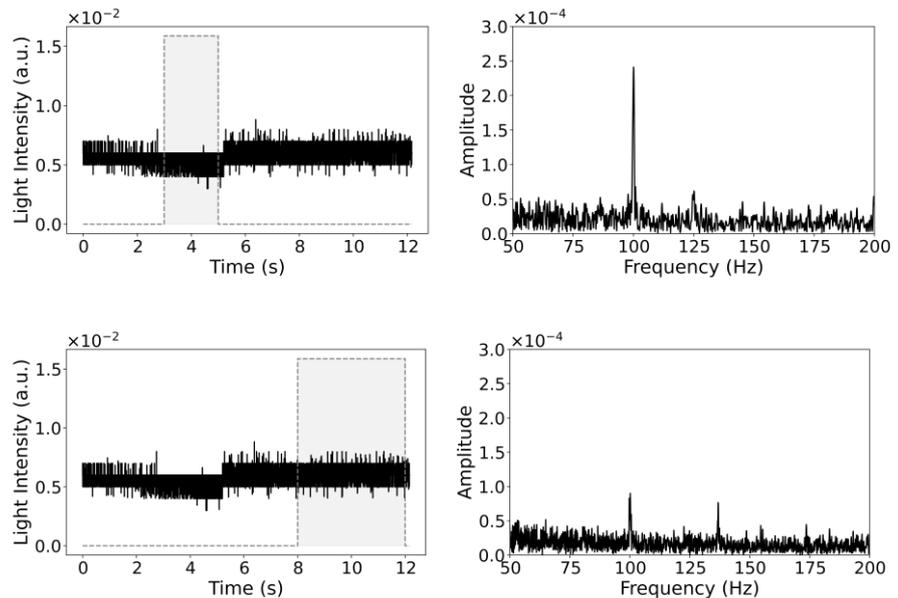

Fig. 3 Same as in Fig.2, here with the ambient light switched on (top row) and with both the ambient light and the loudspeaker switched on (bottom row), whose peaks in the frequency domain (at 100 and 137 Hz, respectively) are shown in the right side of the figure.

iOLab is a highly useful all-in-one data acquisition and analysis device, offering advantages over alternative options. Although this experiment can be performed with other devices, smartphone light sensors usually lack the necessary resolution and appropriate sampling rate to avoid aliasing phenomena, as discussed in the following section. An alternative approach could involve using an appropriate light sensor with an Arduino processor, and analyzing the output with separate FFT code [12].

We also mention that we used a PASCO interferometer/light sensor [14] as a reliable (yet, more expensive) educational apparatus to be compared with the Nikhef instrument. The results are in very good agreement and, as a consequence, we are confident that the cheaper solution of the Nikhef apparatus can be safely adopted in an educational framework.

**3. Measurement of AC power line frequency**

The second experiment in which FFT has a decisive role is devoted to the estimation of the AC power line frequency. This is accomplished once again by using the iOLab sensors but other devices could also be suited to this aim. Since the power line works in alternating current, it is possible to exploit the Faraday law of induction: we built a coil which was placed close to the power cable of a hair dryer and the induced current was amplified and then detected by the iOLab unit as shown in Fig.4. We point out that the FFT process works in accordance with the Shannon-Nyquist theorem, which establishes that the highest frequency that can be determined is one half of the sampling frequency (named also Nyquist frequency). This means that, in the case of the AC signal at 50 Hz, the minimum sampling rate must be 100 Hz. The voltage amplifier interface used by the iOLab works at 200 Hz and this assures that the FFT will be carried out properly. In the results shown in Fig.4 the 50 Hz peak is clearly dominant. As a practical note, we mention that the cord powering the hair dryer needs to be built with two separated, unipolar wires to make possible the magnetic detection of the signal. Moreover, with the coil the frequency of other AC circuits in the environment can also be detected, i.e. one does not need to plug-in the hair dryer to see a signal. It is important to assure that the Shannon-Nyquist theorem can be applied, otherwise an aliasing effect is expected, i.e. the FFT will produce a spectrum which does not contain the actual frequency of the transformed signal. This can be seen, as a practical and educational proficient example of aliasing, in this same experiment carried out in a different way. It is in fact possible to use the iOLab internal magnetic sensor (a Hall probe) in the attempt of directly acquiring the magnetic field generated by the alternating current circulating in the power cord of the hair dryer. The measured field is depicted in Fig.5 along with its FFT frequency spectrum which is now characterized by a peak located at about 32 Hz, whilst the 50 Hz contribution is not present. The reason for this result is that the magnetic field sensor of the iOLab unit has a sampling frequency of about 80 Hz which is smaller than the 100 Hz Nyquist frequency requested to sample without aliasing a 50 Hz signal (the magnetometers in commercial smartphones are also generally not suitable for this specific application, as their sampling frequency typically ranges from 10 Hz to 100 Hz.). It is possible to demonstrate that the FFT output in this case will contain a dominant component whose frequency is given by the difference between the sampling frequency and the actual frequency of the analyzed signal. In our case, the precise sampling frequency of the iOLab during the measurement was 82 Hz which explains the observed peak at 82 Hz-50 Hz=32 Hz, as depicted in Fig.5.

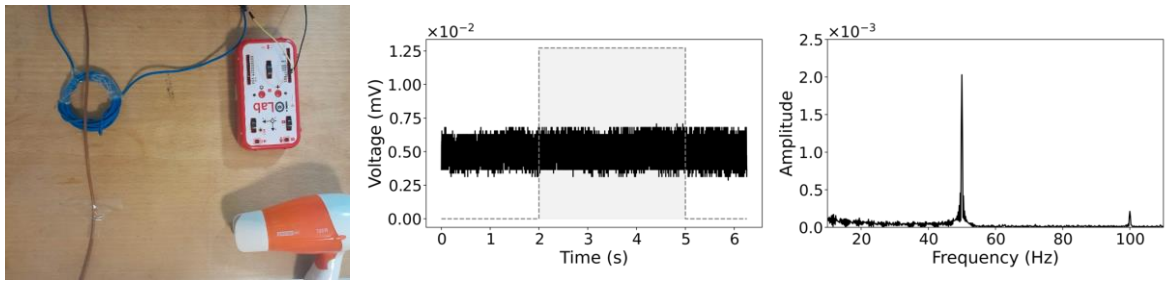

Fig.4 Experimental setup to measure the magnetic field associated with the alternating current powering the hair dryer (left); induced current in the coil (center) and the associated FFT amplitude spectrum showing the 50 Hz peak of the power line (right).

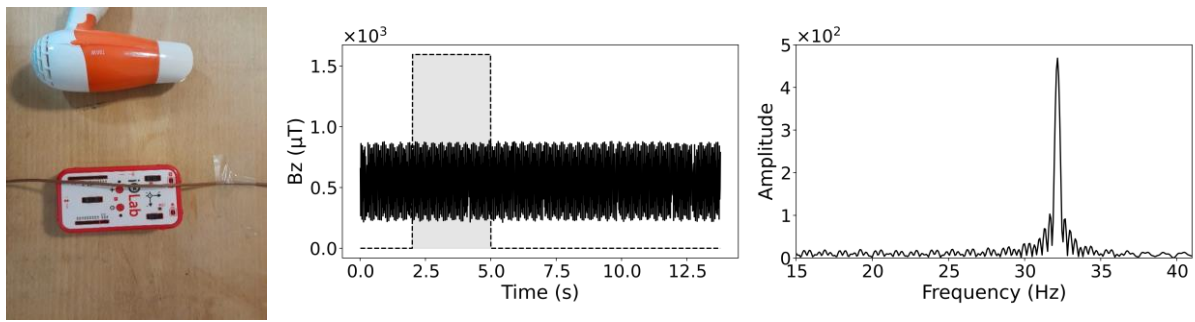

Fig.5 Experimental setup attempting a direct measurement of the magnetic field associated with the electric line powering the hair dryer (left); output of the Hall sensor of the iOLab unit (center) and the associated FFT amplitude spectrum showing the aliased peak at about 32 Hz (right).

**4. Conclusions**

In this work we suggest to exploit digital devices to enrich the exploration and understanding of a very large number of phenomena and physics experiments. The main aim of this approach is related to the possibility, made so simple and immediate thanks to the digital management and treatment of acquired data, to obtain the frequency domain representation of relevant variables in terms of a numerical, real-time FFT. It is true that mathematical technicalities can be addressed properly only at more advanced school levels. Yet, our suggestion is such that, in high school labs for introducing physics studies, math details can be skipped and just the final computed FFT amplitude spectrum can be brought to students' attention as an alternative, graphical and powerful approach for working with periodic signals. These are our conclusions after a series of practical activities with some small groups of teachers and pupils attending the final school years.